\pdfoutput=1

\documentclass[final,5p,times,twocolumn]{elsarticle}

\usepackage{graphicx}

\usepackage{amssymb}
\usepackage{amsmath}
\usepackage{amsthm}

\usepackage{algorithm}
\usepackage{algorithmic}
\usepackage{url}
\usepackage[pdftex,citecolor=blue,linkcolor=blue,linkbordercolor={1 1 1},colorlinks=true,citebordercolor={1 1 1}]{hyperref}

\journal{Information Processing Letters}

\begin{document}

\begin{frontmatter}



\title{A polynomial 3-colorability algorithm with automatic generation of\\ NO 3-colorability (i.e. Co-\textbf{NP}) short proofs}


\author{Jos\'e Antonio Mart\'in H.} \ead{jamartinh@fdi.ucm.es}
\address{Faculty of Computer Science, Complutense University of Madrid, Spain}

\begin{abstract}
In this paper, an algorithm for determining 3-colorability, i.e. the decision problem (YES/NO), in planar graphs is presented. The algorithm, although not exact (it could produce false positives) has two very important features: (i) it has polynomial complexity and (ii) for every ``NO'' answer, a ``short'' proof is generated, which is of much interest since 3-colorability is a \textbf{NP}-complete problem and thus its complementary problem is in Co-\textbf{NP}. Hence the algorithm is exact when it determines that a given planar graph is not 3-colorable since this is verifiable via an automatic generation of short formal proofs (also human-readable).
\end{abstract}


\end{frontmatter}


\section{Introduction}
\label{Introduction}

In graph theory, the case where the task is to find a coloring of the vertex set of some graph, using as few colors as possible, in such a way that if two vertices ($u,v$) are joined by an edge then $u$ and $v$ must receive different colors, is a very well known problem. More specifically, the problem of deciding if a given graph can be colored with at least $k$ colors is called the $k$-colorability problem. For $k=1$ and $k=2$ the vertex coloring problem is trivial. However for $k \geq 3$ in general the problem is \textbf{NP}-complete~\cite{GJ79}. Moreover,  the case of determining if a given planar graph is 3-colorable, i.e., asking for $\chi(G)\leq 3$? remains \textbf{NP}-complete~\cite{Stockmeyer1973} also for some especial classes of planar graphs~\cite{GJ79}.

In this paper an algorithm for determining 3-colorability, i.e., the decision problem (YES/NO), in planar graphs is described. Contrary to common algorithms for solving \textbf{NP}-problems, the presented one produces proofs for negative answers, i.e., proofs of NO 3-colorability. The algorithm is not exact since it could produce false positives or leave the problem as ``UNDETERMINED''. However, it has two very important features:
\begin{enumerate}
  \item It has polynomial complexity in the number ($n$) of vertices.
  \item For every ``NO'' answer, a ``short'' proof is generated, which is of much interest since 3-colorability is a \textbf{NP}-complete problem and thus its complementary problem is in Co-\textbf{NP}~\cite[see,][for related problems on co-\textbf{NP} short proofs]{Boppana1987,Fortnow1988}.
\end{enumerate}
Hence the algorithm is exact when it determines that a given planar graph is not 3-colorable since this can be verified, also with polynomial complexity, via an automatic generation of a human-readable (or compressed, e.g. with a special binary encoding to be automatically verified by a computer program) ``short'' formal proof. Note that it is widely believed that \textbf{NP}$\neq$Co-\textbf{NP}, i.e., that this kind of short proof does not exist in the general case for problems in Co-\textbf{NP}.

Furthermore, the algorithm has been put in the public domain on a web-server\footnote{A polynomial 3-coloring solver with proofs \url{http://totana.dia.fi.upm.es/~jamartinh/ColoringPage.psp}} for testing purposes. The web-server application consist of a server program that can be used to test if an input PLANAR graph submitted trough the web-site in a text file is 3-colorable or not. The web-server accepts planar graphs in the DIMACS\footnote{DIMACS: Center for Discrete Mathematics and Theoretical Computer Science, Rutgers University, \url{http://dimacs.rutgers.edu/}} (.col) standard text format.

After a DIMACS file is uploaded to the server, it attempts to read it as a graph and tries to construct the graph data structure. If the process fails then a message is generated indicating so, and if the process succeeds then the user is notified if the graph is 3-colorable and the total time of computation. Every graph evaluation is archived in a ``files''\footnote{A repository of planar graphs in DIMACS ``.col'' text format is located at \url{http://totana.dia.fi.upm.es/~jamartinh/files/}} directory which is a database of planar graph instances in DIMACS format.

Here we describe in detail the 3-colorability algorithm behind the server.

\section{Preliminary definitions and basic terminology}
\label{sec:1}

Unless we state it otherwise, all graphs in this work are connected and simple (finite, and have
no loops or parallel edges).

Partitioning the set of vertices $V(G)$ of a graph $G$ into
separate classes, in such a way that no two adjacent vertices are
grouped into the same class, is called the vertex graph coloring
problem. In order to distinguish such classes, a set of colors C
is used, and the division into these \emph{(color) classes} is given by a
proper-coloring (we will use here just the single term coloring) $\varphi : V(G)\rightarrow C$, where
$\varphi(u) \neq \varphi(v)$ for all $uv$ belonging to the set
of edges $E(G)$ of $G$. Given a graph vertex coloring problem over
a graph  $G$  with a set of colors $C$, if $C$  has
cardinality  $k$, then $\varphi$ is a \emph{k-coloring} of
$G$. The \emph{Chromatic number} of a graph $\chi(G)$ is the
minimum number of colors necessary to color the vertices of a
graph $G$ in such a way that no two adjacent vertices get colored
with the same color, thus, if $\chi(G) \leq k$ then one says that G
is \emph{k-colorable} (i.e. can be colored with $k$ different colors) and if $\chi(G)= k$ then one says that G is
\emph{k-chromatic}.

An \emph{independent set} (also called \emph{stable set}) $I$  is a set of vertices of $G$ such that there are no edges between any two vertices in $I$.

A \emph{triangle} is a graph or subgraph $T$ consisting of three fully connected vertices (e.g. $\{a,b,c\}$ such that $ab,bc,ac \in E(T)$ )

An important type of graphs are the \emph{planar graphs}, a graph is called planar: if it can be drawn in a plane without edges
crossings.

For more detailed information on graph coloring the reader can see the book of Jensen and Toft on ``Graph Coloring problems''~\cite{JT95}

\section{The Algorithm}
The Algorithm evaluates if a given planar graph $G$ is NOT 3-colorable in polynomial time. Contrary to common algorithms for solving \textbf{NP} decision problems, the presented one produces proofs for negative answers, i.e., proofs of NO 3-colorability. A special feature of this program is that if the answer is NO then \emph{you are sure} that the evaluated planar graph is not 3-colorable since a proof is generated. Note that there is no formal proof that instances returning YES are robust until they return a valid 3-coloring, which is done only in few cases when the algorithm finds a 3-coloring.

\begin{figure}[b]
\hrule $ $\\
\textbf{Preamble}=\\
{\small``\texttt{The graph $G$ is not 3-colorable \\
Proof: (by contradiction) \\
Assume $G$ is 3-colorable, hence it should exist
the possibility of partitioning the vertices of $G$
in three independent sets: $A$, $B$ and $C$.\\
Thus:\\
Select triangle $T=\{a,b,c\}$
and let $A$, $B$, $C$ be three independent sets,
each one containing a different vertex of triangle $T$: \\
$A=[a]$, $B=[b]$, $C=[c]$}''}
\\
\hrule
\caption{The text used as the preamble of the proofs}\label{fig:preamble}
\end{figure}
The generated proof can be divided in three parts:
\begin{enumerate}
  \item Preamble~(Figure~\ref{fig:preamble}).
  \item Rules.
  \item Conclusion~(Figure~\ref{fig:conclusion}).
\end{enumerate}
The \emph{preamble} of the generated proofs is always the same (Figure~\ref{fig:preamble}), expect for the particular values of the vertex labels $a$, $b$ and $c$. The \emph{Rules} are generated on-lime by the heuristics procedures $\rho_1$ and $\rho_2$. Finally, the Conclusion part is always the same (Figure~\ref{fig:conclusion}).

\begin{figure}[tb]
\hrule $ $\\
\textbf{Conclusion}=\\
{\small``\texttt{Thus, does not exist the possibility of partitioning the vertices of $G$ in three independent sets: $A$, $B$ and $C$ (contradiction).\\
Therefore $G$ is not 3-colorable.\\
Q.E.D.}''}
\\
\hrule
\caption{The text used at the final part of the proofs}\label{fig:conclusion}
\end{figure}

\subsection{Main routine: Is-3-colorable(G)}

The main routine of the algorithm, Is-3-colorable(G), consist in: Given a planar-graph $G$ perform the main subroutine, \emph{TestTriangle}$(G,T)$, for every triangle $T=\{a,b,c\}$ of $G$ and returning NO, whenever a proof that $G$ is NOT-3-colorable is found, or YES, otherwise. The routine goes writing a proof $P$ of incremental form, by applying some predefined heuristic rules ($\rho_1$ and $\rho_2$), for each triangle of $G$.

A way of describing the proof ($P$) generation procedure is that the algorithm goes, of incremental form, adding determined vertices to a growing subgraph $P \subseteq G$, in such a way that $xy \in E(P)$ when $xy \in E(G)$, until $P$ becomes a $4$-chromatic graph. The Pseudo-code of the Is-3-colorable(G) routine is shown in Algorithm~\ref{alg:main}.

\begin{algorithm}[h]
   \caption{Is-3-colorable(G)}
   \label{alg:main}
\begin{algorithmic}[1]
   \REQUIRE  A planar graph $G$
   \FORALL{triangle $T$ of $G$}
       \PRINT \textbf{Preamble}
       \STATE $Q \leftarrow$ \emph{TestTriangle}$(G,T)$
       \IF{ $Q$ = \textbf{YES} }
       \RETURN \textbf{YES}
       \ELSIF{ $Q$ = \textbf{NO} }
          \PRINT \textbf{Conclusion}
          \RETURN \textbf{NO}
       \ENDIF
       \STATE clear-screen
   \ENDFOR
   \RETURN \textbf{YES OR UNDETERMINED}
\end{algorithmic}
\end{algorithm}

\subsection{The TestTriangle$(G,T)$ subroutine}
The \emph{TestTriangle} subroutine is designed to construct a proof showing that $G$ is not 3-colorable. The proof method proceeds by contradiction: Assume that $G$ is 3-colorable, then it should exist the possibility of partitioning the vertices of $G$ in three independent sets: $A$, $B$ and $C$. Then by showing that this is not possible the algorithm completes the proof.

The input to the \emph{TestTriangle}$(G,T)$ subroutine is the graph $G$ and a triangle $T=\{a,b,c\}$ of $G$. The first step of the \emph{TestTriangle}$(G,T)$ subroutine is to create three independent sets: $A \leftarrow\{a\}$, $B\leftarrow\{b\}$ and $C\leftarrow\{c\}$ containing the vertices of the triangle $T=\{a,b,c\}$ respectively. The next step is to perform a series of necessary vertex inclusions into $A$, $B$ and $C$ in order to maintain the possibility of partitioning $G$ in three independent sets. These necessary vertex inclusions are determined by heuristics procedures ($\rho_1$ and $\rho_2$), which can be easily extended to incorporate new heuristic vertex inclusion rules. If one of such vertex inclusions fails then it is not possible to partition $G$ in three independent sets and thus the subroutine return NO. A pseudo-code of the \emph{TestTriangle}$(G,T)$ subroutine is shown in Algorithm~\ref{alg:t2}.

\begin{algorithm}[t]
   \caption{\emph{TestTriangle}$(G,T)$ subroutine}
   \label{alg:t2}
\begin{algorithmic}[1]
   \REQUIRE A planar graph $G$ and a triangle $T=\{a,b,c\}$ of $G$
   \STATE $S \leftarrow V(G)$
   \STATE $A \leftarrow\{a\}$, $B\leftarrow\{b\}$, $C\leftarrow\{c\}$
   \STATE $S' \leftarrow S \setminus \{a,b,c\}$
   \WHILE{$|S'|<|S|$}
       \STATE $S=S'$
       \STATE ($Q_1$, $S'$, $A$)  $\leftarrow$ $\rho_1(S',A,G)$
       \STATE ($Q_2$, $S'$, $B$)  $\leftarrow$ $\rho_1(S',B,G)$
       \STATE ($Q_3$, $S'$, $C$)  $\leftarrow$ $\rho_1(S',C,G)$
       \STATE ($Q_4$, $S'$, $C$)  $\leftarrow$ $\rho_2(S',A,B,C,G)$
       \STATE ($Q_5$, $S'$, $B$)  $\leftarrow$ $\rho_2(S',A,C,B,G)$
       \STATE ($Q_6$, $S'$, $A$)  $\leftarrow$ $\rho_2(S',B,C,A,G)$
       \IF{ $Q_i$  = \textbf{NO} for some $i$}
          \RETURN \textbf{NO}
       \ENDIF
       \IF{$S'= \emptyset$}
            \STATE clear-screen
            \PRINT ``{\small\texttt{$G$ is 3-colorable!, a solutions is:}''}
            \PRINT $A$, $B$, $C$
            \RETURN \textbf{YES}
       \ENDIF
   \ENDWHILE
   \RETURN \textbf{UNDETERMINED}
\end{algorithmic}
\end{algorithm}

%

\subsection{The $\rho_1$ and $\rho_2$ subroutines}
The $\rho_1$ and $\rho_2$ subroutines are based on a very simple rules:
\begin{description}
  \item[$\rho_1$:] Given a subset $S$ of vertices of a planar graph $G$ and an independent set $I$ of $G$: if there is a triangle ($\{a,b,c\}$) in $S$ such that $a,b$ are both joined to some element in $I$ then any proper 3-coloring of $G$ should assign vertex $c$ to an independent set not containing $a$ nor $b$, i.e, $I$. Moreover if $c$ is also joined to some element of $I$ then $G$ is not 3-colorable.
  \item[$\rho_2$:] Given a subset $S$ of vertices of $G$, three independent sets $I_a$, $I_b$ and $I_c$ and a planar graph $G$: Every vertex of $S$ having an edge with a vertex of set $I_a$ and a vertex of $I_b$ $(e\{x,b\})$ must be assigned to set $I_c$. Moreover if $c$ is also joined to some element of $I_c$ then $G$ is not 3-colorable.
\end{description}

\begin{algorithm}[h]
   \caption{Subroutine $\rho_1$}
   \label{alg:ro1}
\begin{algorithmic}[1]
   \REQUIRE  A subset $S$ of vertices of $G$, an independent set $I$ and the graph $G$.
      \FORALL{$a,b \in S$ such that $ab\in E(G)$}
          \IF{ $ax \in E(G)$ \textbf{and} $by \in E(G)$; for some $x,y \in I$ }
             \FORALL{$c\in S$ such that $\{a,b,c\}$ is a triangle of $G$}
                   \STATE $S\leftarrow S \setminus \{c\}$
                   \STATE $I\leftarrow I \cup \{c\}$
                   \PRINT {\small``\texttt{Triangle $\{a,b,c\}$ has edges:$e\{a,x\}$ and $e\{b, y\}$ with set $I$ hence vertex $c$ must be assigned to $I$}''}
             \ENDFOR
          \ENDIF
          \IF{ $cz \in E(G)$ for some vertex $z \in I$}
             \PRINT  {\small``\texttt{Every vertex of triangle $\{a,b,c\}$ is joined by an edge to an element in the set $I$}''}
             \RETURN \textbf{NO}
          \ENDIF
   \ENDFOR
   \RETURN \textbf{UNDETERMINED}, $S$, $I$
\end{algorithmic}
\end{algorithm}

\begin{algorithm}[h]
   \caption{Subroutine $\rho_2$}
   \label{alg:ro2}
\begin{algorithmic}[1]
   \REQUIRE  A subset $S$ of vertices of $G$, three independent sets $I_a$, $I_b$ and $I_c$ and the graph $G$.
       \FORALL{x in $S$}
          \IF{ $xa \in E(G)$ for some vertex $a \in I_a$ \textbf{and} $xb \in E(G)$ for some vertex $b \in I_b$}
                \IF{ $xc \in E(G)$ for some vertex $c \in I_c$}
                    \PRINT  {\small``\texttt{Vertex $x$ has an edge with at least, one vertex of set $I_a$ one of $I_b$ and one of $I_c$}''}
                    \RETURN \textbf{NO}
                \ENDIF
                \STATE $S   \leftarrow S   \setminus \{x\}$
                \STATE $I_c \leftarrow I_c \cup \{x\}$
                \PRINT  {\small``\texttt{Vertex $x$ has an edge with one vertex of set $I_a$ $(e\{x,a\})$ and one of $I_b$ $(e\{x,b\})$ hence}''}
                \PRINT  {\small``\texttt{vertex $x$ must be assigned to set $I_c$}''}
          \ENDIF
       \ENDFOR
   \RETURN \textbf{UNDETERMINED}, $S$, $I_c$
\end{algorithmic}
\end{algorithm}
\clearpage
\subsection{Depurating the proof}
Although correct, proof statements (some elements)  introduced by the $\rho_1$ and $\rho_2$ subroutines may be superfluous for the proof due to they are not necessary to maintain the validity of the proof.

This could be the case, for instance, of some vertex inclusions that are then never related with the final part of the proof. For this reason, it is necessary, for clarity and for simplicity purposes, to maintain the generated proof as short and clear as possible with a procedure to analyze and delete all such unnecessary proof statements. We may also recall that the proof generation procedure can be seen as the incremental construction of a 4-chromatic subgraph and thus we can delete all non-critical vertices of the proof $P$.

However, I will keep the presented algorithms as they where described, to maintain its simplicity and clarity, but showing just some required minimal modifications in order to filter the unnecessary proof statements.

First, we can create a proof structure consisting of a hashtable where the keys are the vertices added to each set $A$, $B$, or $C$ and the values are a tuple consisting of:
\begin{itemize}
  \item The actual text $p$ of the proof statements, i.e., {\small\texttt{Triangle $\{a,b,c\}$ has edges:$e\{a,x\}$ and $e\{b, y\}$ with set $I$ hence vertex $c$ must be assigned to $I$}}
  \item A list of necessary immediately precedent vertex inclusions to which the proof statement refers, i.e., $\{x,y\}$
\end{itemize}
And then use a simple procedure (Algorithm~\ref{alg:findvertices}) starting at the last inserted vertex to recover all sufficient and necessary vertices and its respective proof's text line(s).

In this way, clearer and concise proofs can be obtained replacing the ``\textbf{PRINT}'' instructions, in the $\rho_1$ and $\rho_2$ algorithm's (Algorithms~\ref{alg:ro1} and~\ref{alg:ro2}), by a statement such as:
\begin{equation}
p \leftarrow p + \mbox{``original algorithm's proof text''}
\end{equation}
where $p$ is a string containing the actual proof's text.

And then, adding an instruction to fill the proof's hashtable $P$ with the data of the current proof's line:
\begin{equation}
P[v] \leftarrow  \left\{ p,  \{a,b\} \right\},
\end{equation}
where $v$ is the inserted vertex (in the $I$ set), $\{a,b\}$ are the necessary immediately precedent vertex inclusions to which the proof statement refers and $p$ is a text containing the current ``human readable'' proof. 

Then calling $\rho_1$ and $\rho_2$ will be now of the form:
\begin{equation}
(Q_i, P,...)  \leftarrow \rho_i(P,...),
\end{equation}
to make that $P$ persist over calls.

And finally, the Is-3-colorable(G) routine can print out the ``depurated'' proof's text returned by the \emph{DecodeProof}() sub routine just before the \textbf{Conclusion} text is printed out.

\begin{algorithm}[h]
   \caption{DecodeProof($x,P,p,V$)}
   \label{alg:findvertices}
\begin{algorithmic}[1]
   \REQUIRE  a vertex $x$, a proof structure $P$, a proof's text $p$ and a set of vertices $V$
   \FORALL{ related vertices $y$ in $P[x]$ }
       \STATE $(V,p) \leftarrow (V,p) (\cup,+)$ \emph{DecodeProof}($y,P,p,V$)
       \STATE $p \leftarrow  p + P[y].p$
   \ENDFOR
   \RETURN $V,p$
\end{algorithmic}
\end{algorithm}

\subsection{Complexity Analysis}

To determine the computational complexity of the whole algorithm we will start by analyzing from the top routine: Is-3-colorable(G).

The Is-3-colorable routine has complexity of order at most $O(n^3)$ since it explores each triangle and this can be done easily by exploring every combination of three vertices to look for a triangle:

\begin{equation}\label{eq:cc1}
    \mbox{Is-3-colorable} = {n\choose 3} = O(n^3).\mbox{\emph{TestTriangle}}
\end{equation}

The \emph{TestTriangle} function may looks a little bit more difficult since the \textbf{While} loop, but it is also very simple: This function will iterate until the set $S$ becomes empty or, even earlier, when no element is removed from the set $S$ by any of the two subroutines $\rho_1$ and $\rho_2$. Thus it has only three possibilities at every iteration:
\begin{enumerate}
  \item Decrease the number of elements of $S$.
  \item End due the number of elements in $S$ has not decreased.
  \item End due $S$ is empty.
\end{enumerate}

The worst scenario is the third case, that is, when ending due to $S$ is empty since this imply that the loop has been executed \emph{at most} $n$ times. Thus \emph{TestTriangle} has linear complexity $O(n)$.

\begin{equation}\label{eq:cc2}
    \mbox{\emph{TestTriangle}} = {n\choose 1} = O(n).(\rho_1 + \rho_2)
\end{equation}

The $\rho_1$ subroutine performs a loop over every edge in $S$ ($S$ grows linear as it is a subset of $n$) and for every edge another loop is executed for every element in $S$. Thus it has complexity at most: $O(n^3)=O(n^2)O(n)$:

\begin{equation}\label{eq:cc3}
    \rho_1 = {n\choose 2}.{n\choose 1} = O(n^2)O(n)=O(n^3)
\end{equation}

The $\rho_2$ subroutine performs a loop over each element $x$ in $S$ and for each $x$ it looks for a vertex in the set $I_a$, looks for another vertex in $I_b$ and, in the worst case, also for a vertex in $I_c$. All $I_th$ sets are subsets of $n$ so they grow linear w.r.t. $n$. Thus it has complexity at most:  $n\times 3n$ which is  $O(n^2)=O(n)O(n)$:
\begin{equation}\label{eq:cc3}
    \rho_2 = {n\choose 1}.{n\choose 1} = O(n)O(n)=O(n^2)
\end{equation}
Hence the complexity of the Is-3-colorable(G) algorithm is:
\begin{equation}\label{eq:cctotal}
    \mbox{Is-3-colorable}= O(n^3)O(n)\left(O(n^3)+O(n^2)\right)= O(n^7)
\end{equation}

The total complexity of the whole algorithm is analyzed in Algorithm~\ref{alg:bigA}, in an analogous form to the Big-$O$ as an algorithm's definition.
\begin{algorithm}[tb]
   \caption{Run time and grow analysis of the Algorithm}
   \label{alg:bigA}
\begin{algorithmic}[1]
    \FORALL{Is-3-colorable $\rightarrow O(n^3)$}
       \FORALL{\emph{TestTriangle} $\rightarrow O(n)$}
          \STATE $3 \times \rho_1 \rightarrow O(n^3)$
          \STATE $3 \times \rho_2 \rightarrow O(n^2)$
       \ENDFOR       
       $$\mbox{\emph{TestTriangle}} = \left(O(n^3)+O(n^2)\right) O(n) = O(n^4)$$      
   \ENDFOR   
   $$\mbox{Is-3-colorable} = O(n^4) O(n^3) = O(n^7)$$   
\end{algorithmic}
\end{algorithm}

Also, with respect to the complexity of the proof in size terms, we can just show that the size of the proofs grows linear, $O(n)$,in the number $n$ of vertices since every line of the proof does either:
\begin{enumerate}
  \item Insert a vertex into an independent set and eliminates it from the list of available vertices.
  \item Stop the program and return NO with the current proof.
\end{enumerate}
In general, for every NO answer, the algorithm will always insert into the proof less vertices than the number of vertices of the given planar graph and for each inserted vertex there is a constant overhead of text depending on the language. Hence, the generated proofs are really short.

\subsection{A demonstrative example}
\begin{figure}[tb]
  \center
  \includegraphics[scale=0.5]{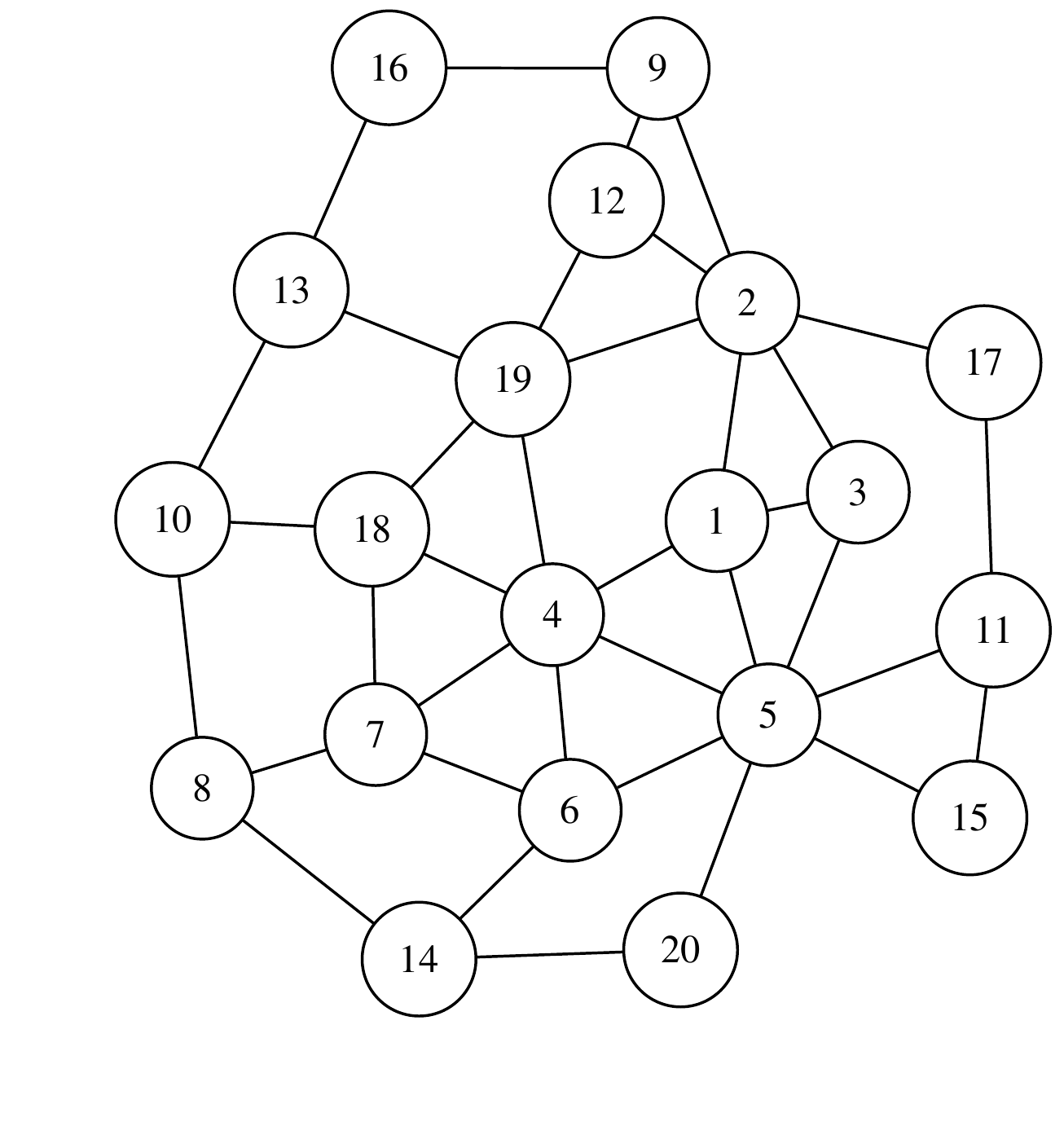}\\
  \caption{The non 4-chromatic planar graph ($G$) used in the demonstrative example}\label{fig:test}
\end{figure}

In order to show an example of the algorithm's output, i.e., a proof of NO-3-colorability, a random planar graph (Figure~\ref{fig:test}) was generated. The graph $G$ has 20 vertices and 36 edges. Figure~\ref{fig:test_result} shows the text of the generated proof.

\begin{figure}[h]
\hrule $ $
\\
{\small\texttt{
The graph $G$ is not 3-colorable.
\\
Proof: (by contradiction)
\\
Assume $G$ is 3-colorable, hence it should exist the possibility of partitioning the vertices of G in three independent sets: $A$, $B$ and $C$
\\
Thus:
\\
Select triangle $T=[1, 2, 3]$
\\
Let $A$, $B$, $C$  be three independent sets, each one containing a different vertex of triangle $T$:
\\
$A=[1]$, $B=[2]$, $C=[3]$
\begin{itemize}
  \item Triangle $[4, 5, 6]$ has edges: $e[1, 4]$ and $e[1, 5]$ with set A hence vertex 6 must be assigned to set A.
  \item Vertex 5 has an edge with one vertex of set $A$ ($e[5, 6]$) and one of $C$ ($e[3, 5])$ hence vertex 5 must be assigned to set $B$.
  \item Triangle $[4, 7, 18]$ has edges: $e[4, 6]$ and $e[6, 7]$ with set $A$ hence vertex 18 must be assigned to set $A$.
  \item Vertex 4 has an edge with one vertex of set $A$ ($e[4, 18]$) and one of $B$ ($e[4, 5]$) hence vertex 4 must be assigned to set $C$.
  \item Vertex 19 has an edge with at least, one vertex of set $A$ ($e[18, 19]$) one of $B$ ($e[2, 19]$) and one of $C$ ($e[4, 19]$).
\end{itemize}
Thus, does not exist the possibility of partitioning the vertices of $G$ in three independent sets: $A$, $B$ and $C$ (contradiction).
\\
Therefore $G$ is not 3-colorable.
\\
Q.E.D.
\\
}}
\hrule
\caption{The output of the algorithm obtained for the test graph $G$ of Figure~\ref{fig:test}}\label{fig:test_result}
\end{figure}


\section{Conclusion and further work}
In this paper, an algorithm for determining 3-colorability in planar graphs has been presented. The algorithm, although not exact, because it could produce false positives, has two very important features: 
\begin{itemize}
  \item It has polynomial complexity in the number ($n$) of vertices. 
  \item For every ``NO'' answer, a ``short'' proof is generated, which is of much interest since 3-colorability is a \textbf{NP}-complete problem and thus its complementary problem is in Co-\textbf{NP}.
\end{itemize}

The algorithm has been tested on a big set of randomly generated planar graphs with different sizes and ration between vertices/edges obtaining exact results for each instance, however the algorithm is not exact since it will not cover all the planar graphs.

A web-server has been designed in order that the interested researchers can test specific planar graph instances over the proposed algorithm and observe the results. Indeed the server seems to be useful for particular research purposed to obtain proofs (i.e. certificates) of NO 3-colorability of some graphs representing particular problems. 

Unfortunately, the coloring-server has not been used so much and it remains as an unknown resource. I have the expectation that this publication helps to disseminate the server and the algorithm, since the intention is to receive comments in order to improve the algorithm by progressively incorporating more heuristics and more web resources such as automatic translation of instances of other \textbf{NP}-complete problems to 3-coloring obtaining thus proofs for a wide range of Co-\textbf{NP} problems.

\bibliographystyle{elsarticle-num}
\bibliography{colorref}

\begin{thebibliography}{1}
\expandafter\ifx\csname url\endcsname\relax
  \def\url#1{\texttt{#1}}\fi
\expandafter\ifx\csname urlprefix\endcsname\relax\def\urlprefix{URL }\fi
\expandafter\ifx\csname href\endcsname\relax
  \def\href#1#2{#2} \def\path#1{#1}\fi

\bibitem{GJ79}
M.~R. Garey, D.~S. Jhonson, Computers and Intractability, A Guide to the Theory
  of NP-Completeness, W.H. Freeman and Co., San Francisco, 1979.

\bibitem{Stockmeyer1973}
L.~Stockmeyer, Planar 3-colorability is polynomial complete, SIGACT News 5
  (1973) 19--25.
\newblock \href {http://dx.doi.org/10.1145/1008293.1008294}
  {\path{doi:10.1145/1008293.1008294}}.

\bibitem{Boppana1987}
R.~B. Boppana, J.~Hastad, S.~Zachos, Does co-np have short interactive proofs?,
  Information Processing Letters 25~(2) (1987) 127--132.
\newblock \href {http://dx.doi.org/10.1016/0020-0190(87)90232-8}
  {\path{doi:10.1016/0020-0190(87)90232-8}}.

\bibitem{Fortnow1988}
L.~Fortnow, M.~Sipser, Are there interactive proofs for {co-NP} languages?,
  Information Processing Letters 28 (1988) 249--251.

\bibitem{JT95}
T.~R. Jensen, B.~Toft, Graph coloring problems, Wiley-Interscience Series in
  Discrete Mathematics and Optimization, John Wiley \& Sons, Chichester-New
  York-Brisbane-Toronto-Singapore, 1995.

\end{thebibliography}

\end{document}